# Three-dimensional Random Voronoi Tessellations: From Cubic Crystal Lattices to Poisson Point Processes

**Valerio Lucarini**

Department of Physics, University of Bologna, Viale Berti Pichat 6/2, 40127, Bologna, Italy
Istituto Nazionale di Fisica Nucleare – Sezione di Bologna, Viale Berti Pichat 6/2, 40127, Bologna, Italy
Email: lucarini@adgb.df.unibo.it

**Abstract**

We perturb the simple cubic (SC), body-centered cubic (BCC), and face-centered cubic (FCC) structures with a spatial Gaussian noise whose adimensional strength is controlled by the parameter $\alpha$ and analyze the statistical properties of the cells of the resulting Voronoi tessellations. We concentrate on topological properties, such as the number of faces, and on metric properties, such as the area and the volume. The topological properties of the Voronoi tessellations of the SC and FCC crystals are unstable with respect to the introduction of noise, because the corresponding polyhedra are geometrically degenerate, whereas the tessellation of the BCC crystal is topologically stable even against noise of small but finite intensity. Whereas the average volume of the cells is the intensity parameter of the system and does not depend on the noise, the average area of the cells has a rather interesting behavior with respect to noise intensity. For weak noise, the mean area of the Voronoi tessellations corresponding to perturbed BCC and FCC perturbed increases quadratically with the noise intensity. In the case of perturbed SCC crystals, there is an optimal amount of noise that minimizes the mean area of the cells. Already for a moderate amount of noise ($\alpha > 0.5$), the statistical properties of the three perturbed tessellations are indistinguishable, and for intense noise ($\alpha > 2$), results converge to those of the Poisson-Voronoi tessellation. Notably, 2-parameter gamma distributions constitute an excellent model for the empirical pdf of all considered topological and metric properties. By analyzing jointly the statistical properties of the area and of the volume of the cells, we discover that not only the area and the volume of the cells fluctuate, but so does their shape too. The shape of a cell can be measured with the isoperimetric quotient, indicating how spherical the cell is. The Voronoi tessellations of the BCC and of the FCC structures result to be local maxima for the isoperimetric quotient among space-filling tessellations. In the BCC case, this suggests a weaker form of the Kelvin conjecture, which has recently been disproved. Due to the fluctuations of the shape of the cells, anomalous scalings with exponents larger than 3/2 are observed between the area and the volumes of the cells for all cases considered, and, except for the FCC structure, also for infinitesimal noise. In the Poisson-Voronoi limit, the exponent is ~1.67. As the number of faces is heavily correlated with the sphericity of the cells (cells with more faces are bulkier), the anomalous scaling is heavily reduced when we perform power law fits separately on cells with a specific number of faces.

# 1. Introduction

A Voronoi tessellation (Voronoi 1907; 1908) is a partitioning of an Euclidean *N*-dimensional space $\Omega$ defined in terms of a given discrete set of points $X \subset \Omega$. For almost any point $a \in \Omega$ there is one specific point $x \in X$ which is closest to *a*. Some point *a* may be equally distant from two or more points of *X*. If *X* contains only two points, $x_1$ and $x_2$, then the set of all points with the same distance from $x_1$ and $x_2$ is a hyperplane, which has codimension 1. The hyperplane bisects perpendicularly the segment from $x_1$ and $x_2$. In general, the set of all points closer to a point $x_i \in X$ than to any other point $x_j \neq x_i$, $x_j \in X$ is the interior of a convex (*N*-1)-polytope usually called the Voronoi cell for $x_i$. The set of the (*N*-1)-polytopes $\Pi_i$, each corresponding to - and containing - one point $x_i \in X$, is the Voronoi tessellation corresponding to *X*. Extensions to the case of non-Euclidean spaces have also been presented (Isokawa 2000; Okabe et al. 2000).

Since Voronoi tessellations boil up to being optimal partitionings of the space resulting from a set of *generating points*, they have long been considered for applications in several research areas, such as telecommunications (Sortais et al. 2007), biology (Finney 1975), astronomy (Icke 1996), forestry (Barrett 1997) atomic physics (Goede et al. 1997), metallurgy (Weaire et al. 1986), polymer science (Dotera 1999), materials science (Bennett et al. 1986), and biophysics (Soyer et al. 2000) In a geophysical context, Voronoi tessellations have been widely used to analyze spatially distributed observational or model output data (Tsai et al. 2004; Lucarini et al. 2007, Lucarini et al 2008). In condensed matter physics, the Voronoi cell of the lattice point of a crystal is known as the Wigner-Seitz cell, whereas the Voronoi cell of the reciprocal lattice point is the Brillouin zone (Bassani and Pastori Parravicini 1975; Ashcroft and Mermin 1976). Voronoi tessellations have been used for performing structure analysis for crystalline solids and supercooled liquids (Tsumuraya et al. 1993; Yu et al., 2005), for detecting glass transitions (Hentschel et al. 2007), for emphasizing the geometrical effects underlying the vibrations in the glass (Luchnikov et al., 2000), and for performing detailed and efficient electronic calculations (Averill and Painter 1989; Rapcewicz et al. 1998). For a review of the theory and applications of Voronoi tessellations, see Aurenhammer (1991) and Okabe et al. (2000).

As the theoretical results on the statistical properties of general N-dimensional Voronoi tessellations are still relatively limited, direct numerical simulation constitute a crucial investigative approach. At computational level, the evaluation of the Voronoi tessellation of a given discrete set of points *X* is not a trivial task, and the definition of optimal procedures is ongoing and involves various scientific communities (Bowyer 1981; Watson 1981; Tanemura et al. 1983; Barber et al. 1996; Han and Bray 2006). The specific and relevant problem of computing the geometric characteristics of Poisson-Voronoi tessellations has been the subject of intense theoretical and computational effort. Poisson-Voronoi tessellations are obtained starting from for a random set of points *X* generated as output of a homogeneous (in the considered space) Poisson point process. This problem has a great relevance at practical level because it corresponds, *e.g.*, to studying crystal aggregates with random nucleation sites and uniform growth rates. Exact results concerning the mean statistical properties of the interface area, inner area, number of vertices, etc. of the Poisson-Voronoi tessellations have been obtained for Euclidean spaces (Meijering 1953; Christ et al., 1982; Drouffe and Itzykson 1984; Calka 2003; Hilhorst 2006); an especially impressive account is given by Finch (2003). Several computational studies on 2D and 3D spaces have found results in agreement with the theoretical findings, and, moreover, have shown that both 2-parameter (Kumar et al. 1992) and 3-parameter (Hinde and Miles 1980) gamma distributions fit up to a high degree of accuracy the empirical pdfs of several geometrical characteristics of the cells (Zhu et al. 2001; Tanemura 2003). The *ab-initio* derivation of the pdf of the geometrical properties of Poisson-Voronoi tessellations have not been yet obtained, except in asymptotic regimes (Hilhorst 2005), which are, surprisingly, not compatible with the gamma distributions family.

In a previous paper (Lucarini 2008), we have analyzed in a rather general framework the statistical properties of Voronoi tessellation of the Euclidean 2D plane. In particular, we have followed the transition from regular triangular, square and hexagonal honeycomb Voronoi tessellations to those of the Poisson-Voronoi case, thus analyzing in a common framework symmetry-break processes and the approach to uniformly random distributions of tessellation-generating points, which is basically realized when the typical displacement becomes larger than the lattice unit vector. This analysis has been accomplished by a Monte Carlo analysis of tessellations generated by points whose regular positions are perturbed through a Gaussian noise. The symmetry break induced by the introduction of noise destroys the triangular and square tessellation, which are structurally unstable, whereas the honeycomb hexagonal tessellation is stable also for small but finite noise. Moreover, independently of the unperturbed structure, for all noise intensities (including infinitesimal), hexagons constitute the most common class of cells and the ensemble mean of the cells area and perimeter restricted to the hexagonal cells coincides with the full ensemble mean, which reinforces the idea that hexagons can be taken as *generic* polygons in 2D Voronoi tessellations.

The reasons why the regular hexagonal tessellation has such peculiar properties of robustness relies on the fact that it is optimal both in terms of perimeter-to-area ratio and in terms of *cost* (see Newman 1982; Du and Wang 2005). The *extremal* properties of such a tessellation are clearly highlighted by Karch et al. (2006), where it is noted that a Gibbs system of repulsive charges in 2D arranges spontaneously for low temperatures (freezes) as a regular hexagonal crystal.

Moreover, a regular hexagonal structure has been found for the Voronoi tessellation built from the spontaneously arranged lattice of hot spots (strongest upward motion of hot fluid) of the Rayleigh-Bènard convective cells, with the compensating downward motion of cooled fluid concentrated on the sides of the Voronoi cells (Rapaport 2006).

In this paper we want to extend the analysis performed in Lucarini (2008) to the 3D case, which is probably of wider applicative interest. We consider three cubic crystals covering the 3D Euclidean space, namely the simple cubic (SC), the face-centered cubic (FCC) and the body-centered cubic (BCC) lattices (Bassani and Pastori Parravicini 1975). The corresponding space-filling Voronoi cells of such crystals are the cube, the rhombic dodecahedron, and the truncated octahedron. The cubic crystal system is one of the most common crystal systems found in elemental metals, and naturally occurring crystals and minerals. These crystals feature extraordinary geometrical properties:

- the cube is the only space-filling regular solid;
- the FCC (together with the Hexagonal Close Packed structure)features the largest possible packing fraction - the 1611 Kepler's conjecture has been recently proved by Hales (2005);
- the Voronoi cell of BCC has been conjectured by Kelvin in 1887 as being the space-filling cell with the smallest surface to volume ratio, and only recently a very cumbersome counter-example has been given by Wearie and Phelan (1994); moreover, the truncated octahedron is conjectured to be have the lowest *cost* among all 3D space-filling cells (see Du and Wang 2005).

Because of its low density, basically due to the low packing faction, the SC system has a high energy structure and is rare in nature, and it is found only in the alpha-form of Po. The BCC is a more compact system and have a low energy structure, is therefore more common in nature. Examples of BCC structures include Fe, Cr, W, and Nb. Finally, thanks to its extremal properties in terms of packing fraction and the resulting high density, FCC crystals are fairly common and specific examples include Pb, Al, Cu, Au and Ag.

The extremal properties of the BCC structure can basically be interpreted as the fact that the corresponding Voronoi cell defines a natural *discrete* mathematical measure, and imply that truncated octahedra constitute an optimal tool for achieving data compression (Entezari et al. 2008). Another outstanding property is that the Voronoi cell of the BCC structure, as opposed to SC and FCC, is topologically stable with respect to infinitesimal perturbations to the position of the lattice points (Troadec et al. 1998).

Using an ensemble-based approach, we study the break-up of the symmetry of the SC, BCC and FCC systems and of their corresponding Voronoi tessellations by stochastically perturbing with a space-homogeneous Gaussian noise of parametrically controlled strength the positions of the lattice points $x_i$, and quantitatively evaluating how the statistical properties of the geometrical characteristics of the resulting 3D Voronoi cells change. The strength of considered perturbation ranges up to the point where typical displacements become larger than the lattice unit vector, which basically leads to the limiting case of the Poisson-Voronoi process. Therefore, our work joins on the analysis of Voronoi cells resulting from infinitesimal (namely, small) perturbations to regular cubic lattices to fully random tessellations.

Our paper is organized as follows. In section 2 we discuss some general properties of the Voronoi tessellations considered, describe the methodology of work and the set of numerical experiments performed. In section 3 we show our results. In section 4 we present our conclusions and perspectives for future work.

## 2. Theoretical and Computational Issues

### *General Properties of Voronoi tessellations*

We consider a random point process characterized by a spatially homogeneous coarse-grained intensity $\rho_0$, such that the expectation value of the number of points $x_i$ in a generic region $\Gamma \in \mathbb{R}^3$ is $\rho_0|\Gamma|$, where $|\Gamma|$ is the Lebesgue measure of $\Gamma$, whereas the fluctuations in the number of points are $\approx \sqrt{\rho_0|\Gamma|}$. If $\rho_0|\Gamma| \gg 1$, we are in the thermodynamic limit and boundary effects are negligible, so that the number of cells of the Voronoi tessellation resulting from the set of points $x_i$ and contained inside $\Gamma$ is $N_V \approx \rho_0|\Gamma|$.

In this paper we consider perfect crystals, crystals with random dislocations, and sparse points resulting from a spatially homogeneous Poisson process. Perfect crystals are obtained when the probability distribution function (pdf) of the random point process can be expressed as a sum of Dirac masses obeying a discrete translational symmetry. Crystals with random dislocations are periodical is a statistical sense since the pdf of the point-process is a non-singular Lebesgue measurable function obeying discrete translational symmetry. Finally, the pdf of the homogeneous Poisson point process is constant in space.

Using scaling arguments (Lucarini 2008), one obtains that in all cases considered the statistical properties of the Voronoi tessellation are intensive, so that the point density $\rho_0$ can be scaled to unity, or, alternatively, the domain may can be scaled to the Cartesian cube $\Gamma_1 = [0,1] \otimes [0,1] \otimes [0,1]$. We will stick to the second approach. We then define $\mu(Y)$ ($\sigma(Y)$) as the mean value (standard deviation) of the variable $Y$ over the $N_V$ cells for the single realization of the random process, whereas the expression $\langle E \rangle$ ($\delta[E]$), indicates the ensemble mean (standard deviation) of the random variable $E$. We have that $\langle \mu(V) \rangle, \langle \sigma(V) \rangle \propto \rho_0^{-1}$, where $V$ is the volume of the Voronoi cell, $\langle \mu(A) \rangle, \langle \sigma(A) \rangle \propto \rho_0^{-2/3}$, where $A$ is the surface area of the Voronoi cell, and $\langle \mu(P) \rangle, \langle \sigma(P) \rangle \propto \rho_0^{-1/3}$, where $P$ is the total perimeter of the cell. The proportionality constants depend of the specific random point process considered. Therefore, by multiplying the ensemble mean estimators of the mean and standard deviation of the various geometrical properties of the Voronoi cells times the appropriate power of $\rho_0$, we obtain universal functions.

Going to the topological properties of the cells, we remind that, since each Voronoi cell is convex, its vertices ($v$), edges ($e$), and faces ($f$) are connected by the simplified Euler-Poincare formula for 3D polyhedra $v - e + f = 2$. Moreover, in a generic solid vertices are trivalent (i.e. given by the intersection of three edges), so that $e = 3/2 v$, which implies that $f = 1/2 v + 2$, so that the knowledge of the number of vertices of a cell provides a rather complete information about the polyhedron. Another general result is that in each cell the average number ($n$) of sides of each face is $n = 6v/(v+2) = 6 - 12/f < 6$, which marks a clear difference with respect to the plane case, where the Euler theorem applies (Lucarini 2008).

## *Exact Results*

We first consider the perfect SC, BCC, and FCC cubic crystals having a total of $\rho_0$ lattice points per unit volume and $\rho_0$ corresponding Voronoi cells in $\Gamma_1$. Therefore, the length of the side of the cubes of the SC, BCC, and FCC crystals are $\rho_0^{-1/3}$, $2^{1/3}\rho_0^{-1/3}$, and $4^{1/3}\rho_0^{-1/3}$, respectively. Basic Euclidean geometry allows us to fully analyze these structures. The cells of the Voronoi tessellation of the SC crystal are cubes (having 12 edges, 6 faces, 8 trivalent vertices) of side length $\rho_0^{-1/3}$ and total surface area $A = 6\rho_0^{-2/3}$. The cells of the Voronoi tessellation of the BCC crystal are truncated octahedra (having 36 edges, 14 faces, 24 trivalent vertices) of side length $2^{-7/6}\rho_0^{-1/3} \approx 0.4454\rho_0^{-1/3}$ and total surface area $A = 3(1 + 2\sqrt{3})2^{-4/3}\rho_0^{-2/3} \approx 5.3147\rho_0^{-2/3}$. The cells of the Voronoi tessellation of the FCC crystal are rhombic dodecahedra (having 24 edges, 12 faces, 6 trivalent vertices, 8 tetravalent vertices) of side length $2^{-4/3}\sqrt{3}\rho_0^{-1/3} \approx 0.6874\rho_0^{-1/3}$ and total surface area $A = 3 \cdot 2^{5/6}\rho_0^{-2/3} \approx 5.3454\rho_0^{-2/3}$. The standard isoperimetric quotient $Q = 36\pi V^2/S^3$, which measures the in non-dimensional units the surface-to-volume ratio of a solid ($Q = 1$ for a sphere), is 0.5236, 0.7534, and 0.7405 for the SC, BCC, and FCC structures, respectively.

On the other end of the "spectrum of randomness", exact results have been obtained on Poisson-Voronoi tessellations using rather cumbersome analytical tools. We report some of the results discussed by Finch (2003):

- the average number of vertices is $\langle\mu(v)\rangle = \frac{96\pi^2}{35} \approx 27.0709$ and its standard deviation is $\langle\sigma(v)\rangle \approx 6.6708$; exploiting the Euler-Poincare relation plus the genericity property, we obtain $\langle\mu(e)\rangle = 3/2\langle\mu(v)\rangle$, $\langle\mu(f)\rangle = 1/2\langle\mu(v)\rangle + 2$, $\langle\sigma(e)\rangle = 3/2\langle\sigma(v)\rangle$, and $\langle\sigma(f)\rangle = 1/2\langle\sigma(v)\rangle$;

- the average surface area is $\langle\mu(A)\rangle = \left(\frac{256\pi}{3}\right)^{1/3}\Gamma\left(\frac{5}{3}\right)\rho_0^{-2/3} \approx 5.8209\rho_0^{-2/3}$ (with $\Gamma(\bullet)$ here indicating the usual Gamma function), and its standard deviation is $\langle\sigma(A)\rangle \approx 04804\rho_0^{-2/3}$;

- the average volume is, by definition, $\langle\mu(V)\rangle = \rho_0^{-1}$, whereas its standard deviation is $\langle\sigma(V)\rangle \approx 0.4231\rho_0^{-1}$.

*Simulations*

For the SC, BCC, and FCC lattices, we introduce a symmetry-breaking 3D-homogeneous ε-Gaussian noise, which randomizes the position of each of the points $x_i$ about its deterministic position with a spatial variance $|\varepsilon^2|$. By defining $|\varepsilon^2| = \alpha^2/\rho_0 = \alpha^2 l_Q^2$, thus expressing the mean squared displacement as a fraction $\alpha^2$ of the inverse of the density of points, which is the natural squared length scale $l_Q^2$. The parameter $\alpha^2$ can be loosely interpreted as a normalized temperature of the lattice. Note that in all cases, when ensembles are considered, the distribution of the $x_i$ is still periodic. The statistical analysis is performed over 100-members ensembles of Voronoi tessellations generated for all values of $\alpha$ ranging from 0 to 2 with step 0.01, plus additional values aimed at checking the weak- and high-noise limits. Another set of simulations is performed by computing an ensemble of 100 Poisson-Voronoi tessellations generated starting from a set of uniformly randomly distributed $\rho_0$ points per unit volume.

The actual simulations are performed by applying, within a customized routine, the MATLAB7.0® functions voronoin.m and convhulln.m, which implement the algorithm introduced by Barber et al. (1996), to a set of points $x_i$ having coarse grained density $\rho_0 = 100000$ and generated according to the considered random process. The function voronoin.m associates to each point the vertices of the corresponding Voronoi cell and its volume, whereas the function convhulln.m is used to generate the convex hull of the cell.

Note that the convex hull is given in terms of 2-simplices, *i.e.* triangles. Whereas this information is sufficient for computing the total surface area of the cell, an additional step is needed in order to define the topological properties of the cell. In fact, in order to determine the actual number of faces of the cell and define exactly what polygon each face is, we need to explore whether neighbouring simplices are coplanar, and thus constitute higher order polygons. This is accomplished by computing the unit vector $\hat{i}_k$ orthogonal to each simplex $s_k$ and computing the matrix of the scalar products $\langle \hat{i}_j, \hat{i}_k \rangle$ for all the simplices of the cell. When the scalar products $\langle \hat{i}_p, \hat{i}_q \rangle$, $m \leq p,q \leq m+n-1$ of unit vectors orthogonal to $n$ neighbouring simplices $s_k$ $m \leq k \leq m+n-1$ are close to 1 - within a specified tolerance $\xi$, corresponding to a tolerance of about $\sqrt{2\xi}$ in the angle between the unit vector - we have that $\bigcup_{k=m}^{m+n-1} s_k$ is a polygon with $n+2$ sides. We have consistently verified that choosing any tolerance smaller than $\xi = 10^{-8}$ we obtain basically the same results.

Other values of $\rho_0$ - smaller and larger than $\rho_0 = 100000$ - have been used in order to check the previously described scaling laws, which are found to be precisely verified in all numerical experiments. The benefit of using such a large value of $\rho_0$ relies in the fac6t that fluctuations in the ensembles are quite small. Tessellation has been performed starting from points $x_i$ belonging to the square $[-0.1, 1.1] \otimes [-0.1, 1.1] \otimes [-0.1, 1.1] \supset \Gamma_1 = [0,1] \otimes [0,1] \otimes [0,1]$, but only the cells belonging to $\Gamma_1$ have been considered for evaluating the statistical properties, in order to basically avoid $\rho_0$ depletion in the case of large values of α due to one-step Brownian diffusion of the points nearby the boundaries.

## 3. From Regular to Random Lattices

By definition, if α = 0 we are in the deterministic case of SC, BCC, and FCC lattices. We study how the geometrical properties of the Voronoi cells change with α, covering the whole range going from the symmetry break, occurring when α becomes positive, up to the progressively more and more uniform distribution of $x_i$, obtained when α is large with respect to 1 and the pdfs of nearby points $x_i$ overlap more and more significantly.

*Faces, Edges, Vertices*

When spatial noise is present in the system, the resulting Voronoi cells are generic polyhedra, so that degenerate quadrivalent vertices, such as those present in the rhombic dodecahedron (Troadec et al. 1998) are removed with probability 1. Therefore, we expect that $\langle\mu(e)\rangle = 3/2\langle\mu(v)\rangle$, $\langle\mu(f)\rangle = 1/2\langle\mu(v)\rangle + 2$, $\langle\sigma(e)\rangle = 3/2\langle\sigma(v)\rangle$, and $\langle\sigma(f)\rangle = 1/2\langle\sigma(v)\rangle$. These relations have been verified up to a very high degree of accuracy in our simulations, so that, in order to describe the topology of the cell, it is sufficient to present the statistical properties of just one among *e*, *f*, and *v*. In Figure 1 we present our results relative to the number of faces of the Voronoi cells.

In the case of the SC and FCC crystals, the introduction of a minimal amount of symmetry-breaking noise induces a transition the statistics of $\mu(f)$ and $\sigma(f)$, since $\langle\mu(f)\rangle$ and $\langle\sigma(f)\rangle$ are discontinuous in $\alpha = 0$. In the SC case, the average number of faces jumps from 8 to over 16, whereas, as discussed by Troadec et al. (1998), the disappearance of the quadrivalent vertices in the rhombic dodecahedron case causes an increase of two units (up to exactly 14) in the average number of faces. Near $\alpha = 0$, for both SC and FCC perturbed crystals $\langle\mu(f)\rangle$ depends linearly on $\alpha$ as $\langle\mu(f)\rangle \approx \langle\mu(f)\rangle\big|_{\alpha=0+} + \gamma\alpha$, where by $\langle\bullet\rangle\big|_{\alpha=0+}$ we mean the limit for infinitesimal noise of the quantity $\langle\bullet\rangle$. The proportionality constant has opposite sign in the two cases, with $\gamma \approx -1.5$ for the SC and $\gamma \approx 1$ for the FCC case.

Moreover, the introduction of noise generates the sudden appearance of a finite standard deviation in the number of faces in each cell $\langle\sigma(f)\rangle\big|_{\alpha=0+}$, which is larger for SC crystals. In the case of FCC crystals, Troadec et al. (1998) propose a theoretical value $\langle\sigma(f)\rangle\big|_{\alpha=0+} = \sqrt{4/3}$, whereas our numerical estimate is about 10% lower, actually in good agreement with the numerical results presented by Troadec et al. in the same paper. Somewhat surprisingly, the $\langle\sigma(f)\rangle$ is almost constant for a finite range near $\alpha = 0$ ranging up to about $\alpha \approx 0.3$ for the SC crystal and $\alpha \approx 0.2$ for the FCC crystal, thus defining an intrinsic width of the distribution of faces for a – well-defined - "weakly perturbed" state. Thus, we find, in the case of the perturbed FCC crystals, the range of applicability of the weak noise linear perturbation analysis by Troadec et al. (1998).

When considering the BCC crystal, the impact of introducing noise in the position of the points $x_i$ is rather different. Results are also shown in Fig. 1. Infinitesimal noise does not effect at all the tessellation, in the sense that all Voronoi cells are 14-faceted (as in the unperturbed state). Moreover, even finite-size noise basically does not distort cells in such a way that other polyhedra are created,. We have not observed – also going to higher densities - any non-14 faceted polyhedron for up to $\alpha \approx 0.1$ in any member of the ensemble, so that $\langle\mu(f)\rangle=14$ and $\langle\sigma(f)\rangle=0$ in a finite range. However, since the Gaussian noise we are using induces for each point $x_i$ a distribution with – an unrealistic- non-compact support, in principle it is possible to have outliers that, at local level, can distort heavily the tessellation, so that we should interpret this result as that finding non 14-faceted cells is highly – in some sense, exponentially - unlikely.

In Figure 1 the Poisson-Voronoi limiting case is indicated; our simulations provide results in perfect agreement with the analytical by Finch (2003). For $\alpha>1$ the value of $\langle\mu(f)\rangle$ and $\langle\sigma(f)\rangle$ of the Voronoi tessellations of the three perturbed crystals asymptotically converge to what resulting from the Poisson-Voronoi tessellation, as expected. We should note, though, that in the 2D case the asymptotic convergence has been shown to be much slower (Lucarini 2008), so that spatial noise in 3D seems to mix things up much more efficiently. Similarly to the 2D case, the perturbed tessellations are statistically undistinguishable – especially those resulting from the BCC and FCC distorted lattices – well before converging to the Poisson-Voronoi case, thus pointing at some general behavior.

Additional statistical properties of the distribution of the number of faces in the Voronoi tessellation need to be mentioned. The mode of the distribution is quite interesting since the number of faces is, obviously, integer. In the FC and BCC cases, up to $\alpha\approx 0.3$ the mode is 14, whereas for larger values of $\alpha$ 15-faceted polyhedra are the most common ones. Also the Voronoi tessellations of medium-to – highly perturbed SC crystals are dominated by 15-faceted polyhedra, wheras 16-faceted polyhedra dominate up to $\alpha\approx 0.25$.

We also observe that for all non-singular cases - $\alpha>0$ for SC and FCC crystals, $\alpha>0.1$ for BCC crystals -, and *a fortiori* for the Poisson-Voronoi case, the distribution of faces can be represented to a very high degree of precision with a 2-parameter gamma distribution:

$$g(x|k,\theta) = N_V x^{k-1} \frac{\exp[-x/\theta]}{\theta^k \Gamma(k)} \qquad (1),$$

where $\Gamma(k)$ is the usual gamma function and $N_V \approx \rho_0$ is, by definition, the normalization factor. We remind that $\mu(f)=k(f)\theta(f)$ and $\sigma(f)^2=k(f)\theta(f)^2$, so that all information regarding $k(f)$ and $\theta(f)$ can be deduced from Figure 1. In particular, in the Poisson-Voronoi case, our results are in excellent agreement with those of Tanemura (2003). As a side note, we mention that in such a limit case, we observe cells with number of faces ranging from 6 to 36.

Since, as previously discussed, the inverse of number of faces $y=1/f$ is related to the average number $n$ of sides per face in each Voronoi cell as $n = 6 - 12/f = 6 - 12y$, expression (1) can in principle be used also for studying the statistical properties of $n$ and for finding explicitly, with a simple change of variable, the $n$ distribution (or at least a statistical model for it), which is not a 2-parameter gamma distribution. Using expression (1), we have that:

$$\langle\mu(n)\rangle = 6 - 12\langle\mu(y)\rangle = 6 - 12\langle\mu(1/f)\rangle = 6 - 12/(k-1)\theta < 6 - 12/k\theta = 6 - 12/\langle\mu(f)\rangle, \qquad (3)$$

In the Poisson-Voronoi limit, Finch (2003) proposes:

$$\langle\mu(n)\rangle = 6 - 12\langle\mu(1/f)\rangle = 6 - 12/\langle\mu(f)\rangle = 144\pi^2/(24\pi^2 + 35) \approx 5.22. \qquad (4)$$

This results seems to be wrong, as direct numerical simulations give $\langle\mu(n)\rangle = 6 - 12\langle\mu(1/f)\rangle \approx 5.19$. The mistake seems to derive from the fact that the mean of the inverse of the number of faces is different from the inverse of the mean. If, instead, we use expression (3) for computing $\langle\mu(n)\rangle$, thus using the 2-parameter gamma model in the Poisson-Voronoi case, we obtain an almost exact result. The same applies also for all considered values of $\alpha$ and for the three perturbed crystalline structures considered in this study. This further reinforces the idea that 2-parameter gamma family pdfs should be thought as excellent statistical models for the distributions of number of faces in general 3D Voronoi tessellation.

*Area and Volume of the cells*

The statistical properties of the area and of the volume of the Voronoi tessellations of the perturbed cubic crystals have a less pathological behaviour with respect to what previously described when noise is turned on, as all properties are continuous and differentiable in $\alpha = 0$. Still, some rather interesting features can be observed.

As mentioned above, the ensemble mean of the mean volume of the cells is set to $\rho_0^{-1}$ in all cases, so that we discuss the properties of the ensemble mean $\langle\sigma(V)\rangle$, shown in Figure 2. We first observe that for all cubic structures the standard deviation converges to zero with vanishing noise, thus meaning that small variations in the position in the lattice points do not create dramatic rearrangements in the cells when their volumes are considered. Moreover, for $\alpha < 0.3$, a well-defined linear behaviour $\langle\sigma(V)\rangle \approx \alpha X$ is observed for all SC, BCC, and FCC structures. The proportionality constant X is not distinguishable between the BCC and the FCC perturbed crystals, and actually $X \approx 1.2\langle\sigma(V)\rangle_{PV}$, where the pedix refers to the asymptotic Poisson-Voronoi value; the SC curve is somewhat steeper near the origin. The three curves become undistinguishable for $\alpha > 0.4$, so when the noise is moderately intense and reticules are still relatively organized. As previously observed, this seems to be a rather general and robust feature. It is also interesting to note that the attainment of the Poisson-Voronoi limit is quite slow, as compared to the case of the statistical properties of the number of faces of the cell, and a comparable agreement is obtained only for $\alpha > 3$ (not shown).

When considering the area of the cells (see Figure 3), further interesting properties can be highlighted. First, the behaviour of the ensemble mean $\langle\sigma(A)\rangle$ is rather similar to what just discussed for $\langle\sigma(V)\rangle$. Nevertheless, in this case the agreement between the three perturbed structure is more precisely verified – the three curves are barely distinguishable for all values of $\alpha$. Moreover, we again observe a linear behaviour like $\langle\sigma(A)\rangle \approx \alpha\langle\sigma(A)\rangle_{PV}$, which suggests that the systems closely "align" towards Poisson-like randomness.for small values of $\alpha$. This closely mirror what observed in the 2D case for the expectation value of the standard deviation of the perimeter and of the area of the Voronoi cells (Lucarini 2008).

The properties of $\langle\mu(A)\rangle$ for the perturbed crystal structures are also shown in Figure 3. A striking feature is that, similarly to what noted in the case of triangular and square tessellations of the plane, there is a specific amount of noise that optimizes the mean surface for the perturbed SC crystals. We see that the mean area of the cells decreases by about 8% when $\alpha$ is increased from 0 to about 0.3, where a (quadratic) minimum is attained. For stronger noise, the mean area of the Voronoi cells of the perturbed SC crystals decreases, and, for $\alpha > 0.5$, the asymptotic value of the Poisson-Voronoi tessellation is reached. In terms of cell surface minimization, the unperturbed cubic tessellation is about 3% worse than the "most random" tessellation.

The dependence of $\langle\mu(A)\rangle$ with respect to $\alpha$ is very interesting also for the perturbed BCC and FCC cubic crystals. In both cases, the mean area increases quadratically (with very similar coefficient) for small values of $\alpha$, which shows that the Voronoi tessellations of the BCC and FCC cubic crystals are local minima for the mean surface in the set of space-filling tessellations. We know that neither the truncated octahedron nor the rhombic dodecahedron are global minima, since (at least) the Weaire-Phelan structure has a smaller surface (Weaire and Phelan 1994). It is reasonable to expect that a similar quadratic increase of the average surface should be observed when perturbing with spatial gaussian noise the crystalline structure corresponding to the Weaire-Phelan cell. For $\alpha > 0.3$, the values of $\langle \mu(A) \rangle$ for perturbed BCC and FCC crystals basically coincide, and for $\alpha > 0.5$ the Poisson-Voronoi limit is reached within a high accuracy.

Also in the case of cells area and volume, for $\alpha > 0$, the empirical pdfs can be fitted very efficiently using 2-parameter gamma distributions, thus confirming what observed in the 2D case (Lucarini 2008).

*Fluctuations, Shape and Anomalous Scaling*
The analysis of the properties of the joint area-volume pdf for the Voronoi cells of the considered tessellations sheds light on the statistics of fluctuations of these quantities.

As discussed above, the area and the volume of the Voronoi cells resulting from a random tessellations cells are highly variable. See Figure 4 for the joint cells area-volume distribution in the case of Poisson-Voronoi tessellation. All considered perturbed crystal structures give qualitatively similar results, but feature, as obvious from the previous discussion, more peaked distributions. In particular, by integrating the 2D distribution along either direction, we obtain a 1D-pdf which, as previously discussed, is closely approximated with a 2-parameter gamma distribution.

A first interesting statistical property where the joint cells area-volume pdf has to be considered is the isoperimetric quotient $Q = 36\pi V^2 / S^3$. When evaluating its expectation value, we have:

$$\langle \mu(Q) \rangle = 36\pi \langle \mu(V^2/A^3) \rangle \neq 36\pi \langle \mu(V) \rangle^2 / \langle \mu(A) \rangle^3 . \qquad (5)$$

This implies that testing the average "sphericity" – which is basically what $Q$ measures - of a random tessellation is a slightly different problem from testing the average surface for a given average volume, which is what Figure 3 refers to, whereas in regular tessellations the two problems are equivalent. In Figure 5 we present our results for the three perturbed crystal structures. The observed behaviour are qualitatively similar to what could be deduced from Figure 3 using the quantity $36\pi \langle \mu(V) \rangle^2 / \langle \mu(A) \rangle^3$ as an estimate for $\langle \mu(Q) \rangle$ - e.g. $\langle \mu(Q) \rangle$ decreases with $\alpha$ for the BCC and FCC perturbed crystals - but significant additional information emerges. We note that the discrepancy between the estimate and the exact result for $\langle \mu(Q) \rangle$ is positive and monotonically increasing with $\alpha$, with a relative error of the order of 10% when the Poisson-Voronoi limit is attained. Similarly to what observed when analyzing the statistical properties of the area of the cells, in this case we have that by optimally tuning the intensity of the noise ($\alpha \approx 0.3$) perturbing the SC crystal, $\langle \mu(Q) \rangle$ reaches a maximum (with a 20% increase from both the Poisson-Voronoi limit and a 25% increase from the regular crystal limits), so that the corresponding cells are on the average "more spherical". It should also be emphasized that, for all perturbed crystals and for all $\alpha > 0$, $Q$ features a notable variability (in the Poisson-Voronoi limit $Q$ values range from 0.15 to 0.77), which implies that the shape of the cells can vary wildly.

An anomalous scaling is observed when fitting a power law between the value of the area and of and the value of the volume of the cells. When attempting a linear fit between the logarithm of the volume and of the area of cells, we find that the best fit of the exponent $\eta$ such that $V \propto A^\eta$ is in all cases larger than 3/2. The values of the best fit for $\eta$ for the perturbed SC, BCC, and FCC crystals are shown in Figure 6 as a function of $\alpha$; we wish to remark that the quality of the fits is very high, with relative uncertainties of the order of at most $10^{-2}$ in all cases. In particular, in the Poisson-Voronoi limit $\eta \approx 1.67$ - black line in Figure 4 - which suggests the occurrence of a 5/3 exponent. It is also remarkable that, as soon as noise is turned on, anomalous scaling is observed in for the SC and BCC cubic crystals. In the SC case, $\eta(\alpha=0_+) \approx 1.9$, the exponent is monotonically decreasing for all values of $\alpha$, and becomes undistinguishable from the Poisson-Voronoi limit for $\alpha > 2$. In the BCC case, as opposed to what one could expect given the structural stability of the crystal, $\eta(\alpha=0_+) \approx 1.57 > 3/2$, which implies that a (modest) anomalous scaling is observed also for infinitesimal noise. The exponent increases for small values of $\alpha$, overshoots the Poisson-Voronoi limit, and for $\alpha > 0.6$ its value basically coincides wit what obtained in the SC case. When considering the FCC perturbed crystal, an anomalous scaling is osberevd for all finite values of noise, but, quite notably, $\eta(\alpha=0_+) = 3/2$, which means that for infinitesimal noise anomalous scaling is not observed. So, in this regard, the FCC crystal seems to be more robust than the BCC one.

The presence of fluctuations in the values of the area and of the volume of the cells may be invoked as an explanation for the observed anomalous callings. Note that this link is not anobvious one: if the space is tessellated with cubes featuring a spectrum of side lengths, we would in all cases observe a 3/2 scaling exponent between areas and volumes of the cells. Therefore, the reason for the anomalous scaling should be related to the fact that cells of various sizes are not geometrically similar (in a statistical sense), which agrees with the fact that the isoperimetric ratio $Q$ has a distinct variability. Nevertheless, the statistical properties of $Q$ do not explain why the FCC crystal – as opposed to the BCC one- does not feature anomalous scaling in the low noise limit, which implies that the shape of its Voronoi cells are more stable than in the BCC case

In order to clarify the impact of "shape" fluctuations in the estimate of the exponent $\eta$, we take advantage of a strategy of investigation commonly adopted for studying Voronoi tessellations, *i.e.* the stratification of the expectation values of the geometrical properties with respect to classes defined by the number of sides of the cells (Zhou et al. 2001, Hilhorst 2005, Lucarini 2008). In the present case, it would be profitable to study quantities such as $\langle \mu(V) \rangle|_f$ and $\langle \mu(A) \rangle|_f$, where the pedix indicates that the averages are performed only on cells with $f$ faces. It is readily observed that, in all cases, both $\langle \mu(V) \rangle|_f$ and $\langle \mu(A) \rangle|_f$ increase with $f$ (not shown), for the basic and intuitive reason that cells with a larger number of faces are typically larger in volume and have larger areas. Moreover, and this is a more interesting point, larger cells are also typically bulkier, so that their isoperimetric $Q$ is larger. An exact result in this sense is that in 2D Poisson-Voronoi tessellations cells with a large number of sides asymptotically converge to circles (Hilhorst 2005). In Figures 7(a), 7(b), and 7(c) we present the expectation value of the isoperimetric quotient $\langle \mu(Q) \rangle|_f$ as a function of $\alpha$ for the three perturbed cubic crystals. We observe that in all cases, for a given value of $\alpha$, $\langle \mu(Q) \rangle|_f$ increases significantly with $f$, thus confirming the geometrical intuition. Moreover, whereas in the BCC and FCC case $\langle \mu(Q) \rangle|_f$ for a given $f$ decreases monotonically with $\alpha$, in the SC perturbed crystal we have an increase of $\langle \mu(Q) \rangle|_f$ with $\alpha$ for all values of $f$ up to $\alpha \approx 0.3$. Therefore, the isoperimetric quotient optimization due to noise observed in Figure 3 for SC perturbed crystals occurs not only on the overall average, but also separately in each class of cells. For $\alpha > 1$ the results of the three perturbed crystals tend to converge to the Poisson-Voronoi limit. At any rate, the main information contained in Figures 7 is that for all perturbed crystal structures:

a) cells are really very variable in terms of shape;
b) the number of faces acts as a good proxy variable for the isoperimetric quotient, and for the shape of the cells.

Item b) suggests us a strategy for cross-checking the relevance of the fluctuations in the shape of the cells in determining the presence of the anomalous exponent $\eta > 3/2$ depicted in Figure 6. We then attempt a power-law fit $V \propto A^\eta$ separately for each class of cells, so that for each value of noise we obtain via a best fit the optimal $\eta_f$ for the subset of cells having $f$ faces. It is amazing to find that for all cases considered $\eta_f$ results to be lower that the corresponding "global" $\eta$, and for most values of $\alpha$ and $f$ we have that $3/2 < \eta_f < 1.6$. Such behaviour is also independent of the typical population of the classes of cells. We do not plot $\eta_f$ as a function of $\alpha$ for the three perturbed crystals, since what is obtained is just a plateau with no structure along the $\alpha$ or $f$ directions. We then have that the $f$-classification removes most of the shape fluctuations within each class, so that anomalous scaling is greatly suppressed.

We believe that

# 4. Summary and Conclusions

In this work we have studied the statistical properties of the 3D Voronoi tessellations generated by points situated on perturbed cubic crystal structures, namely the SC, BCC, and FCC lattices. The perturbations to the perfect crystal structures are obtained by applying Gaussian noise to the positions of each point.. The variance of the position of the points induced by the Gaussian noise is expressed as $|\varepsilon|^2 = \alpha^2/\rho_0^{2/3}$, where $\alpha$ (which is a sort of square root of a normalized temperature) is the control parameter, $\rho_0$ is the number of points – and of Voronoi cells – per unit volume, and $\rho_0^{-1/3}$ is the resulting natural length scale. By increasing $\alpha$, we explore the transition from perfect crystals to less and less regular structures, until the limit of uniformly random distribution of points is attained. Note that, for all values of $\alpha$, the pdf of the lattice points is in all cases periodic. For each value of $\alpha$, we have performed a set of simulations, in order to create an ensemble of Voronoi tessellations in the unit cube, and have computed the statistical properties of the cells.

The first notable result is that the Voronoi cells corresponding to SC and FCC lattices (cube and rhombic dodecahedron, respectively) are not stable in terms of topological structure. As soon as noise is turned on, their degeneracies – to be intended as non-generic properties of their vertices – are removed via a symmetry-break. With the introduction of an infinitesimal amount of noise, the pdf of the distribution of the number of faces per cell becomes smooth, with a finite standard deviation and a biased mean with respect to the zero noise case. Instead, the zero-noise limit of the Voronoi tessellation of the perturbed BCC structure is coincident with the unperturbed case. Surprisingly, the topology of the perfect BCC tessellation is robust also against small but finite noise. For strong noise, the statistical properties of the tessellations of the three perturbed crystals converge to those of the Poisson-Voronoi limit, but, quite notably, the memory of the specific initial unperturbed state is lost already for moderate noise, since the statistical properties of the three perturbed tessellations are indistinguishable already for $\alpha > 0.5$. As opposed to the 2D case, where hexagon is so overwhelmingly the "generic polygon", since hexagons constitute the stable tessellation, are the most common polygons of basically any tessellation, and the average number of sides of the cells is six (Lucarini 2008), in the 3D case we do not have such a dominant topological structure.

Rather interesting features emerge looking at the metric properties of the cells, namely their surface area and their volume. Given the intensive nature of the Voronoi tessellation, the ensemble average of the mean and of the standard deviation of the area and of the volume of the cell scale as $\rho_0^{-2/3}$ and $\rho_0^{-1}$, respectively, so that we can easily obtain universal results by performing simulations with a specific given density $\rho_0$.

For all values of $\alpha$, the standard deviation of the area of the Voronoi cells is basically the same for the three perturbed cubic structures: it features a linear increase for small noise, and then approaches asymptotically the Poisson-Voronoi limit for large values of $\alpha$. Practically the same applies for the standard deviation of the volume of the cells.

Whereas the ensemble average of the mean volume is set to $\rho_0^{-1}$ by definition, the analysis of the properties of the ensemble average of the mean area of the cells is quite insightful. In the case of perturbed BCC and FCC structures, the mean area increases only quadratically with $\alpha$ for weak noise, thus suggesting that the unperturbed Voronoi tessellation are local minima for the interface area. Instead, the mean area of the perturbed SC structure has a more interesting dependence on noise intensity: it decreases up to $\alpha \approx 0.3$, where a local minimum is attained, wheras for larger values of $\alpha$, it increases until reaching the Poisson-Voronoi limit (as for FCC and BCC). So, counter-intuitively, noise can act as an "optimizer" in the case of the perturbed SC structure.

In all cases analyzed, 2-parameter gamma distributions are very effective in describing the empirical pdfs both for discrete variables, such as the number of faces, and for continuous variable, as in the case of the area and volume of the cells. In particular, such an approach has been useful in identifying and explaining a mistake contained in Finch (2003) regarding the ensemble average of the mean number of sides per face in each Voronoi cell.

The presence of strong fluctuations in the area and volume of the Voronoi cells has suggested to check some joint area-volume statistical properties. In particular, we have considered the isoperimetric quotient $Q$, which measures the sphericity of the cells and is a good quantitative measure of the shape of the cells. The ensemble average of the mean value of $Q$ is quadratic maximum for the FCC and BCC crystals for $\alpha = 0$, and, for any value of $\alpha$, the perturbed BCC structure has the largest mean isoperimetric quotient.. The observation that the truncated octahedron is a "large" local maximum for $Q$ for space-filling tessellations suggests a weak re-formulation of the Kelvin conjecture of global optimality of the truncated octahedron, which has been proved false (Weaire and Phelan 1994).

Moreover, when noise in included in the system, $Q$ features in all cases a rather large variability, which suggests that the shape of the cells can vary a lot within a given tessellation.. In particular, it is observed that the number of faces of a cell is a good proxy for its isoperimetric quotient: cells with a larger number of faces are bulkier. The main effect of the fluctuations in the shape of the cell – and not merely of their area and volume - is that when attempting a power law fit $V \propto A^\eta$ between the areas and the volumes of the Voronoi cells, we obtain for all perturbed tessellations and for any intensity of noise an anomalous scaling, i.e. $\eta > 3/2$, with $\eta \approx 1.67$ in the Poisson-Voronoi limit. Such an anomalous scaling is observed also for infinitesimal noise, except in the case of the FCC crystal, whose tessellation, in this sense, is the most stable. A strong evidence of the relevance of the shape fluctuations in determining the anomalous scaling lies in the fact that it is almost suppressed when we classify the cells according to the number of their faces and attempt the power law fit class by class.

This work clearly defines a way of connecting, with a simple parametric control of spatial noise, crystal structure to uniformly random distribution of points. Such a procedure can be in principle applied for describing the "dissolution" of any crystalline structure. Our results emphasize the importance of analyzing in a common framework the topological and the metric properties of the Voronoi tessellations. In particular, it proves that, in order to grasp a better understanding, it is necessary to consider the joint statistical properties of the area and of the volume of the cells, thus going beyond 1D pdf.. This has proved to be especially useful for framing in all generality the properties of optimality of the FCC and BCC structures.

Several open questions call for answers. The relationship between fluctuations in the shape of the cells and the anomalous scaling, as well as the connection between the number of faces of the cells and their average isoperimetric ratio, surely deserve further analyses. It would be important to analyze the impact of spatial noise on other relevant crystalline structures, such as the lattice whose Voronoi cell is the Weaire-Phelan structure, or the Hexagonal Close Packed crystal, which seems to have close correspondences to the FCC structure, also when infinitesimal perturbation to the position of the points are considered (Troadec et al. 1998). Finally, the impact of noise on higher order statistical properties, such those of neighbouring cells (Senthil Kumar and Kumaran 2005) , should be seriously addressed.

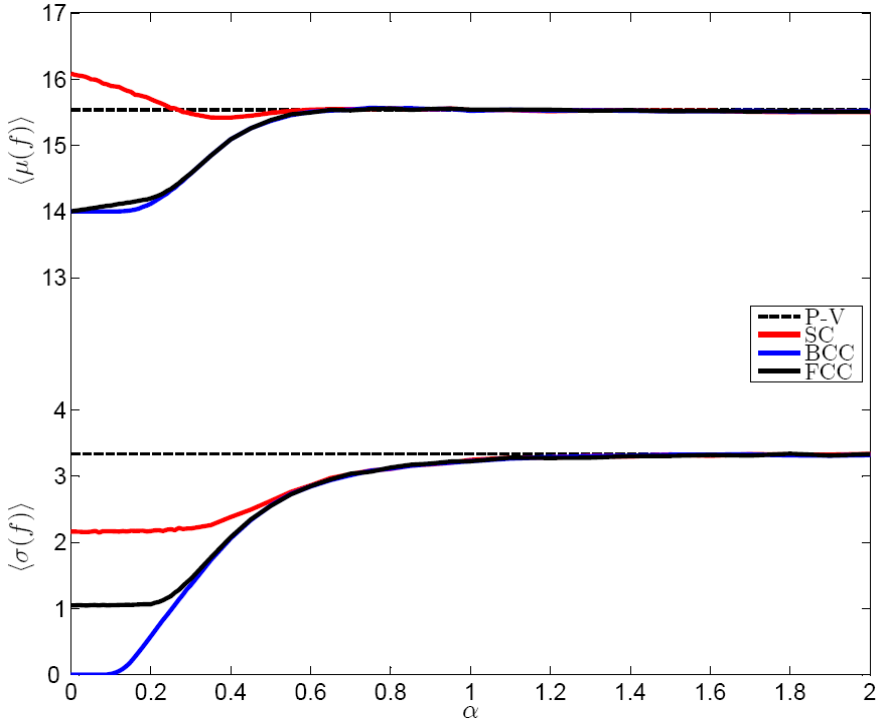

*Figure 1: Ensemble mean of the mean and of the standard deviation of the number of faces (f) of the Voronoi cells for perturbed SC, BCC and FCC cubic crystals. The error bars, whose half-width is twice the standard deviation computed over the ensemble, are too small to be plotted. The Poisson-Voronoi limit is indicated.*

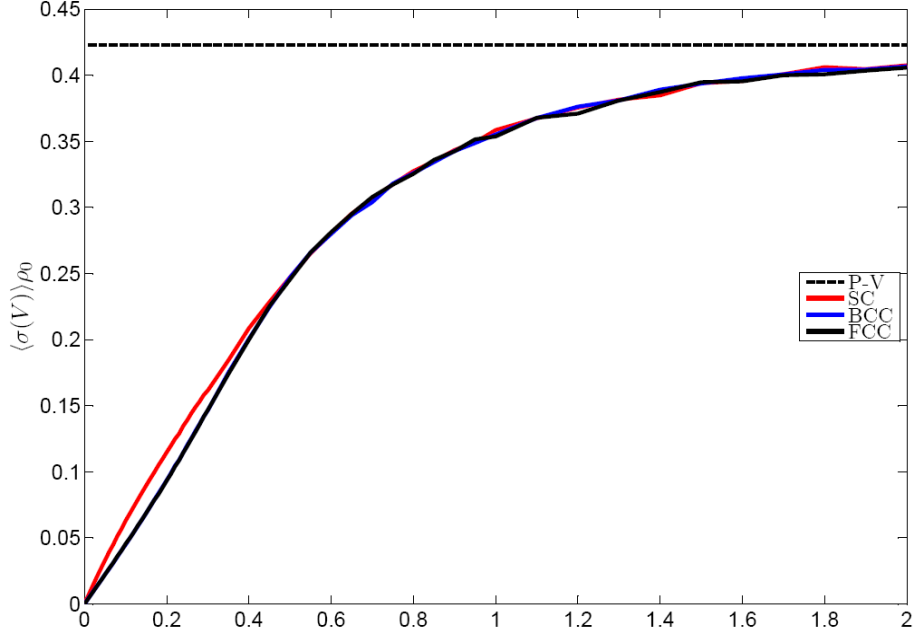

*Figure 2: Ensemble mean of the standard deviation of the volume (V) of the Voronoi cells for perturbed SC, BCC and FCC cubic crystal. The ensemble mean of the mean is set to the inverse of the density. The mean values is Values are multiplied times the appropriate power of the density in order to obtain universal functions. The error bars, whose half-width is twice the standard deviation computed over the ensemble, are too small to be plotted. The Poisson-Voronoi limit is indicated.*

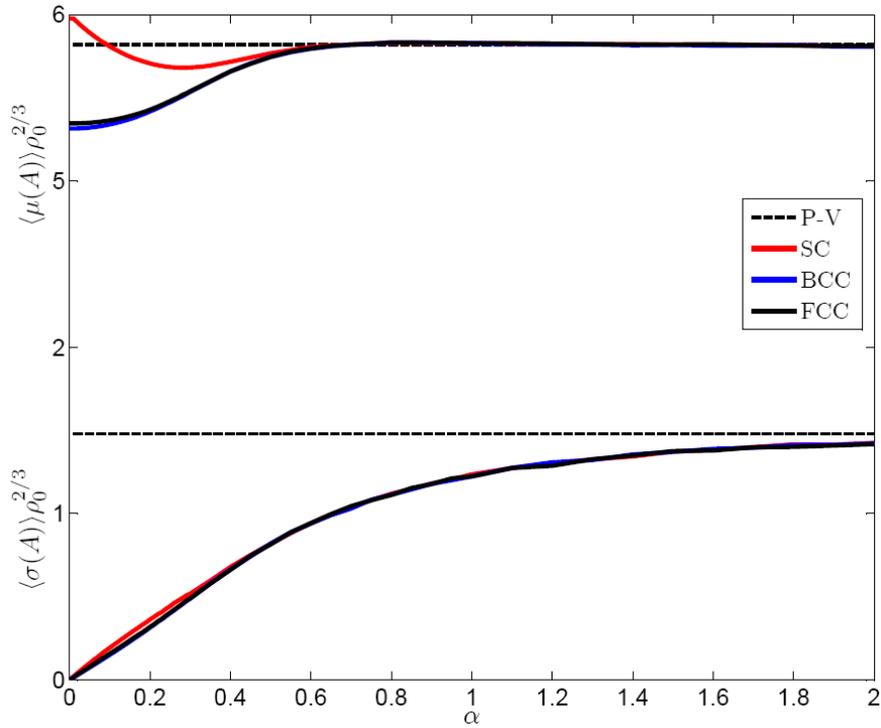

*Figure 3: Ensemble mean of the mean and of the standard deviation of the area (A) of the Voronoi cells for perturbed SC, BCC and FCC cubic crystals. Values are multiplied times the appropriate power of the density in order to obtain universal functions. The error bars, whose half-width is twice the standard deviation computed over the ensemble, are too small to be plotted. The Poisson-Voronoi limit is indicated.*

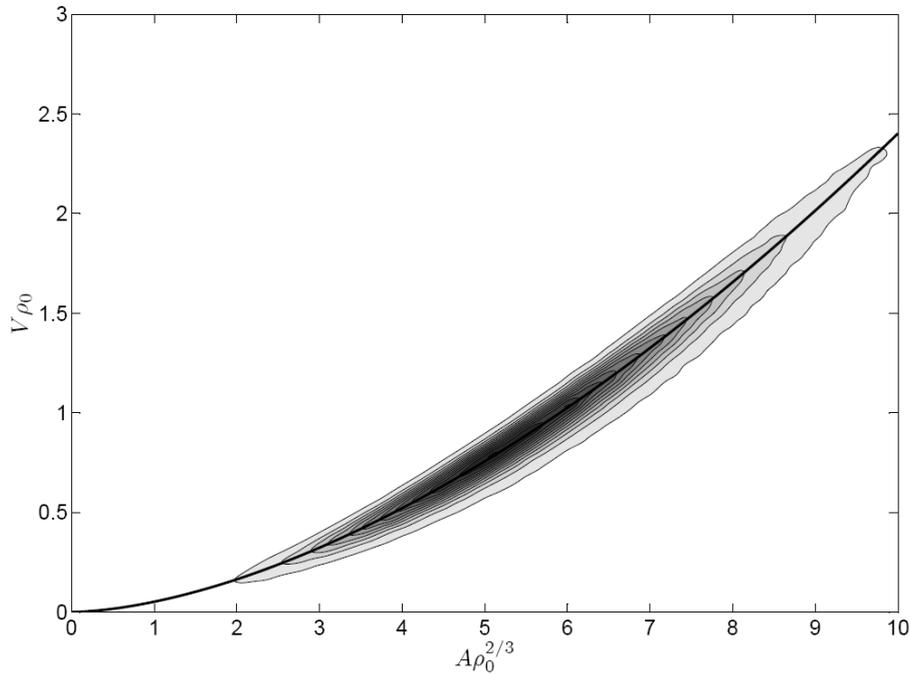

*Figure 4: Joint distribution of the area and of the volume of the Voronoi cells in the Poisson-Voronoi tessellation limit. The black line indicates the best log least squares fit. Details in the text.*

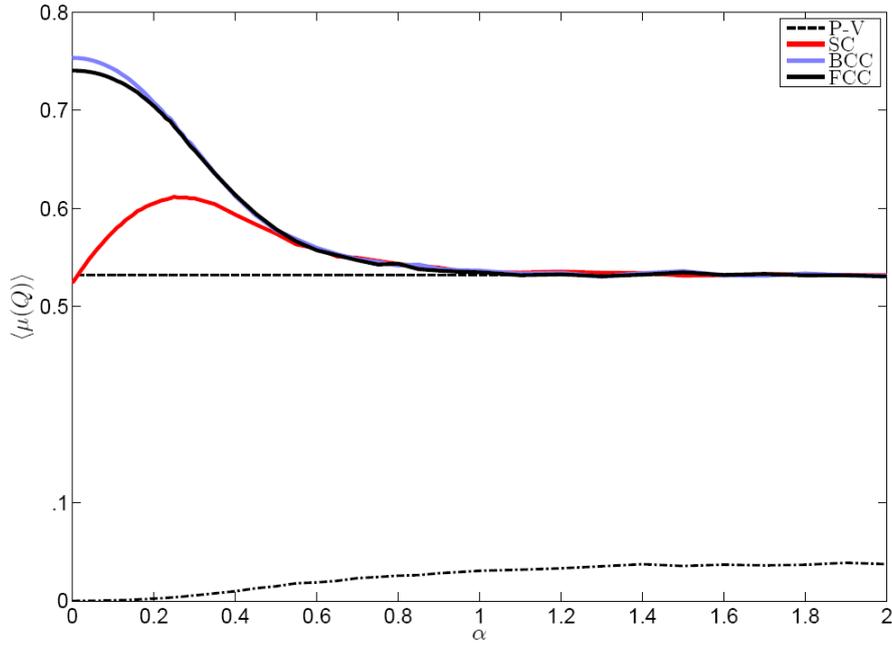

*Figure 5: Ensemble mean of the isoperimetric ratio $Q = 36\pi V^2/A^3$ of the Voronoi cells for perturbed SC, BCC and FCC cubic crystals. The impact of cross-correlations between A and V is given by $\langle \mu(Q) \rangle - 36\pi \langle \mu(V) \rangle^2 / \langle \mu(A) \rangle^3$, which is basically the same for the three perturbed lattices and is plotted with a dashed-dotted line. Details in the text.*

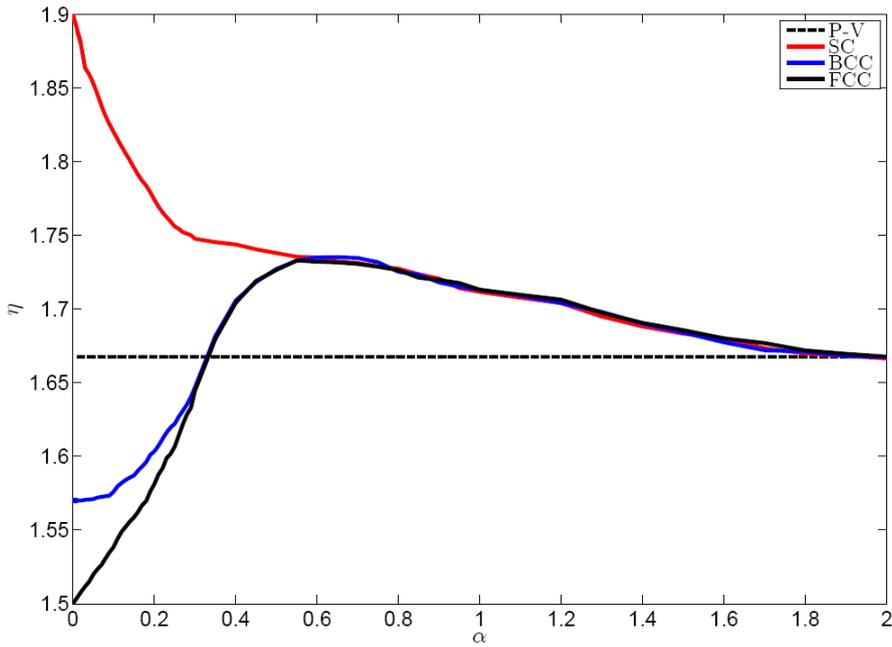

*Figure 6: Scaling exponent $\eta$ fitting the power-law relation $V \propto A^\eta$ for the Voronoi cells of perturbed SC, BCC and FCC cubic crystals. The presence of an anomalous scaling ($\eta \neq 3/2$) due to the fluctuations in the shape of the cells is apparent. Details in the text.*

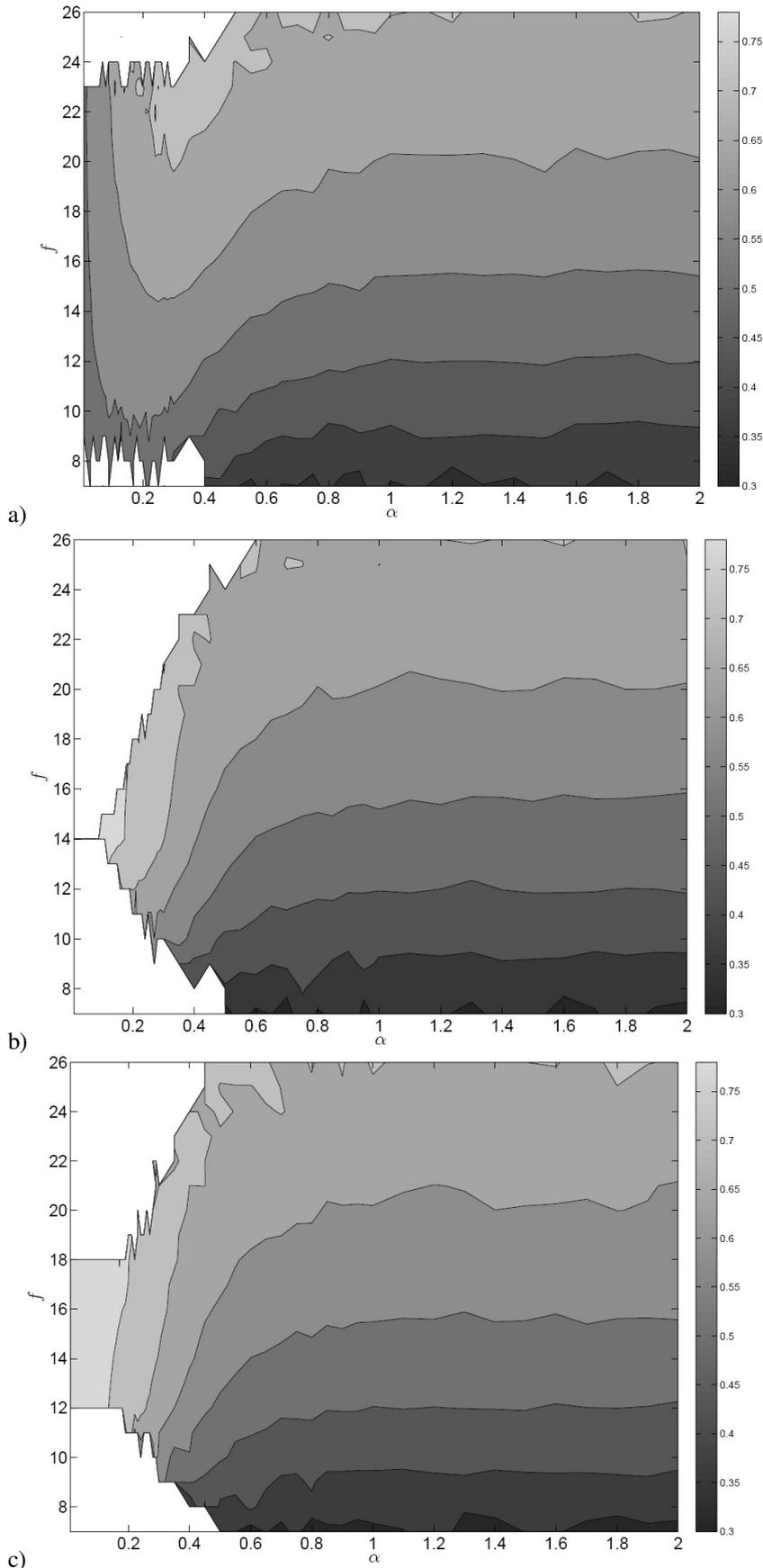

*Figure 7: Ensemble mean of the isoperimetric ratio $\langle \mu(Q) \rangle|_f$ of the Voronoi cells for perturbed SC (a), BCC (b) and FCC(c) cubic crystals, where averages are taken over cells having f faces.. A white shading indicates that the corresponding ensemble is empty. Cells with a larger number of faces are typically more bulky.*

# Bibliography

Ashcroft, N. W. and Mermin, N. D., Solid State Physics, Saunders, 1976.

Aurenhammer F. (1991). Voronoi Diagrams - A Survey of a Fundamental Geometric Data Structure. ACM Computing Surveys, 23, 345-405.

Averill F. W., Painter G. S. (1989), Pseudospherical integration scheme for electronic-structure calculations, Phys. Rev. B 39, 8115

Barber C. B., Dobkin D. P., and Huhdanpaa H.T. (1996). The Quickhull Algorithm for Convex Hulls, ACM Transactions on Mathematical Software 22, 469-483

Barrett T. M. (1997). Voronoi tessellation methods to delineate harvest units for spatial forest planning, Can. J. For. Res. 27(6): 903–910.

Bassani F. and Pastori-Parravicini G., Electronic States and Optical Transitions in Solids, Pergamon Press, Oxford, 1975.

Bennett, L. H., Kuriyama, M., Long, G. G., Melamud, M., Watson, R. E., and Weinert, M. (1986). Local atomic environments in periodic and aperiodic Al-Mn alloys, Phys. Rev. B 34, 8270-8272.

Bowyer A. (1981). Computing Dirichlet tessellations, The Computer Journal 1981 24:162-166.

Calka P. (2003). Precise formulae for the distributions of the principal geometric characteristics of the typical cells of a two-dimensional Poisson Voronoi tessellation and a Poisson line process, Advances in Applied Probability 35, 551-562.

Christ, N. H., Friedberg, R., and Lee, T. D. (1982). Random lattice field theory: General formulation. Nuclear Physics B 202, 89 – 125.

Desch C. H. (1919). The solidification of metals from the liquid state, J. Inst. Metals, 22, 241.

Dotera T. (1999). Cell Crystals: Kelvin's Polyhedra in Block Copolymer Melts, Phys. Rev. Lett. 82, 105–108.

Drouffe J. M. and Itzykson C. (1984). Random geometry and the statistics of two-dimensional cells, Nucl. Phys. B 235, 45-53

Du Q., Wang D. (2005). The Optimal Centroidal Voronoi Tessellations and the Gersho's Conjecture in the Three Dimensional Space, Computers and Mathematics with Applications 49, 1355-1373

Entezari A., Van De Ville D., Möller T. (2008), Practical Box Splines for Reconstruction on the Body Centered Cubic Lattice," IEEE Transactions on Visualization and Computer Graphics 14, 313-328.

Finch S. R., Mathematical Constants, Cambridge University Press, Cambridge, 2003.

Finney, J. L. (1975). Volume occupation, environment and. accessibility in proteins. The problem of the protein surface. J. Mol. Biol. 96, 721–732.

Goede A., Preissner R., and Frömmel C. (1997). Voronoi cell: New method for allocation of space among atoms: Elimination of avoidable errors in calculation of atomic volume and density, J. of Comp. Chem. 18 1113-1118.

Hales T. C. (2005). A Proof of the Kepler Conjecture. Ann. Math. 162, 1065-1185

Han D. and Bray M. (2006). Automated Thiessen polygon generation, Water Resour. Res., 42, W11502, doi:10.1029/2005WR004365.

Hentschel H. G. E., Ilyin V., Makedonska N., Procaccia I., Schupper N., (2007). Statistical mechanics of the glass transition as revealed by a Voronoi tessellation, Phys. Rev. E **75**, 50404(R) (2007)

Hilhorst H. J. (2005). Asymptotic statistics of the n-sided planar Poisson–Voronoi cell: I. Exact results J. Stat. Mech. (2005) P09005   doi:10.1088/1742-5468/2005/09/P09005.

Hinde A. L. and Miles R. E. (1980). Monte Carlo estimates of the distributions of the random polygons of the Voronoi tessellation with respect to a Poisson process. Journal of Statistical Comptutation and Simulation, 10, 205–223.

Icke V (1996). Particles, space and time, Astrophys. Space Sci., 244, 293-312.

Karch R., Neumann M., Neumann F., Ullrich R., Neumüller J., Schreiner W. (2006). A Gibbs point field model for the spatial pattern of coronary capillaries, Physica A 369, 599-611

Kumar, S., Kurtz, S. K., Banavar, J. R. and Sharma, M. G. (1992). Properties of a three-dimensional Poisson-Voronoi tessellation: a Monte Carlo study. Journal of Statistical Physics, 67, 523–551.

Isokawa Y., (2000). Poisson-Voronoi tessellations in three-dimensional hyperbolic spaces, Adv. Appl. Probl. 32, 648-662

Lewis F. T., (1928). The correlation between cell division and the shapes and sizes of prismatic cells in the epidermis of Cucumis, Anat. Rec., 38. 341-376.

Lucarini, V., Danihlik E., Kriegerova I., and Speranza A. (2007). Does the Danube exist? Versions of reality given by various regional climate models and climatological data sets, J. Geophys. Res., 112, D13103, doi:10.1029/2006JD008360.

Lucarini V., Danihlik R., Kriegerova I., and Speranza A. (2008). Hydrological Cycle in the Danube basin in present-day and XXII century simulations by IPCCAR4 global climate models, J. Geophys. Res. 113, D09107, doi:10.1029/2007JD009167

Luchnikov V. A., Medvedev N. N., Naberukhin Yu. I., Schober H. R (2000). Voronoi-Delaunay analysis of normal modes in a simple model glass, Phys. Rev. B **62**, 3181 (2000)

Meijering, J. L., (1953). Interface area, edge length, and number of vertices in crystal aggregates with random nucleation: Phillips Research Reports, Philips Res. Rep., 8, 270-290.

Newman D. (1982), The Hexagon Theorem, IEEE Trans. Inform. Theory 28, 129-137

Okabe, A., Boots B., Sugihara K., and Chiu S. N. (2000). Spatial Tessellations - Concepts and Applications of Voronoi Diagrams. 2nd edition. John Wiley, 2000.

Rapaport D. C. (2006). Hexagonal convection patterns in atomistically simulated fluids, Phys. Rev. E 73, 025301.

Rapcewicz K., Chen B., Yakobson B., Bernholc J., (1998), Consistent methodology for calculating surface and interface energies, Phys. Rev. B **57**, 007281

Senthil Kumar V. and Kumaran V. (2005). Voronoi neighbor statistics of hard-disks and hard-spheres. J. Chem. Phys. 123, 074502.

Sortais M., Hermann S., and Wolisz A. (2007). Analytical Investigation of Intersection-Based Range-Free Localization Information Gain. In: Proc. of European Wireless 2007.

Soyer, A., Chomilier, J., Mornon J.P., Jullien, R. and Sadoc, J.F. (2000). Voronoi tessellation reveals the condensed matter character of folded proteins. Phys Rev Lett, 85, 3532-3535.

Tanemura, M., Ogawa, T., and Ogita, N. (1983). A new algorithm for three-dimensional Voronoi tessellation. Journal of Computational Physics, 51, 191 – 207.

Tanemura M. (2003). Statistical distributions of Poisson-Voronoi cells in two and three Dimensions, Forma 18, 221-247.

Troadec J. P., Gervois A., Oger L. (1998), Statistics of Voronoi cells of slightly perturbed face-centered cubic and hexagonal close-packed lattices, Europhy. Lett. 42, 167-172.

Tsai F. T.-C., Sun N.-Z., Yeh W. W.-G (2004). Geophysical parameterization and parameter structure identification using natural neighbors in groundwater inverse problems, J. Hydrology 308, 269-283.

Tsumuraya K., Ishibashi K., Kusunoki K. (1993), Statistics of Voronoi polyhedra in a model silicon glass Phys. Rev. B **47**, 8552.

Voronoi G. (1907). Nouvelles applications des paramètres continus à la théorie des formes quadratiques. Premier Mémoire: Sur quelques propriétées des formes quadritiques positives parfaites. J. Reine Angew. Math. 133:97-178.

Voronoi G. (1908). Nouvelles Applications des Parametres Continus a la Theorie des Formes Quadratiques. Duesieme Memoire: Recherches sur les Paralloderes Primitifs, J. Reine Angew. Math. 134:198-287.

Watson D. F. (1981). Computing the n-dimensional tessellation with application to Voronoi polytopes, The Computer Journal 1981 24:167-172.

Weaire D., Kermode J.P., and Wejchert J. (1986). On the distribution of cell areas in a Voronoi network. Phil. Mag. B, 53, L101–L105.

Weaire, D. and Phelan, R. (1994) A Counter-Example to Kelvin's Conjecture on Minimal Surfaces. Philos. Mag. Let. 69, 107-110.

Yu D.-Q., Chen M., Han X.-J. (2005) Structure analysis methods for crystalline solids and supercooled liquids, Phys. Rev. E **72**, 051202

Zhu H. X., Thorpe S. M., Windle A. H. (2001). The geometrical properties of irregular two-dimensional Voronoi tessellations, Philosophical Magazine A 81, 2765-2783.